\documentstyle[11pt,newpasp,epsf]{article}
\begin{document}
\title{Dynamical Constraints on Disk Galaxy Formation}
\author{Stacy S. McGaugh}
\affil{Department of Astronomy, University of Maryland}
\begin{abstract}
The rotation curves of disk galaxies exhibit a number of striking
regularities.  The amplitude of the rotation is correlated with
luminosity (Tully-Fisher), the shape of the rotation curve is
well predicted by the luminous mass distribution,
and the magnitude of the mass discrepancy increases
systematically with decreasing centripetal acceleration.
These properties indicate a tight connection between light and
mass, and impose strong constraints on theories of galaxy formation.
\end{abstract}

\keywords{Dark Matter, Galaxies, Galaxy Formation, Modified Dynamics}

\section{Some Important Properties}

There are many systematic properties of the dynamics of spiral galaxies
which galaxy formation theory must explain.  These include (at least)
\begin{enumerate}
\item the luminosity--rotation velocity relation (Tully \& Fisher 1977),
\item the utility of maximum disk in high surface brightness galaxies (van
Albada \& Sancisi 1986; Sellwood 1999),
\item the surface brightness--enclosed mass-to-light ratio relation
(Zwaan et al.\ 1995; McGaugh \& de Blok 1998),
\item the luminosity--rotation curve shape relation (Rubin et al.\ 1985;
Persic \& Salucci 1991),
\item the mass discrepancy--acceleration relation (Sanders 1990;
McGaugh 1999), and
\item the MOND (Milgrom 1983) phenomenology
(Begeman et al.\ 1991; Sanders 1996; de Blok \& McGaugh 1998).
\end{enumerate}
The properties listed above are interrelated, and their meaning
open to interpretation.  I will attempt here to take an empirical approach
illustrating the general requirements imposed on theory by the data.

\section{The Baryonic Mass--Rotation Velocity Relation}

The relation between galaxy luminosity and line width is well known (the
Tully-Fisher relation).  This important property (1) is illustrated in Fig. 1.

Common lore has it that the Tully-Fisher relation is a reflection of an
underlying relation between total mass and rotation velocity.
Presumably, luminosity is closely proportional to total mass.
If this is the case, it makes sense that for gas rich galaxies, a
more appropriate quantity to plot than luminosity would be the sum
of stellar and gaseous mass (Fig. 1).  

\begin{figure}
\plotone{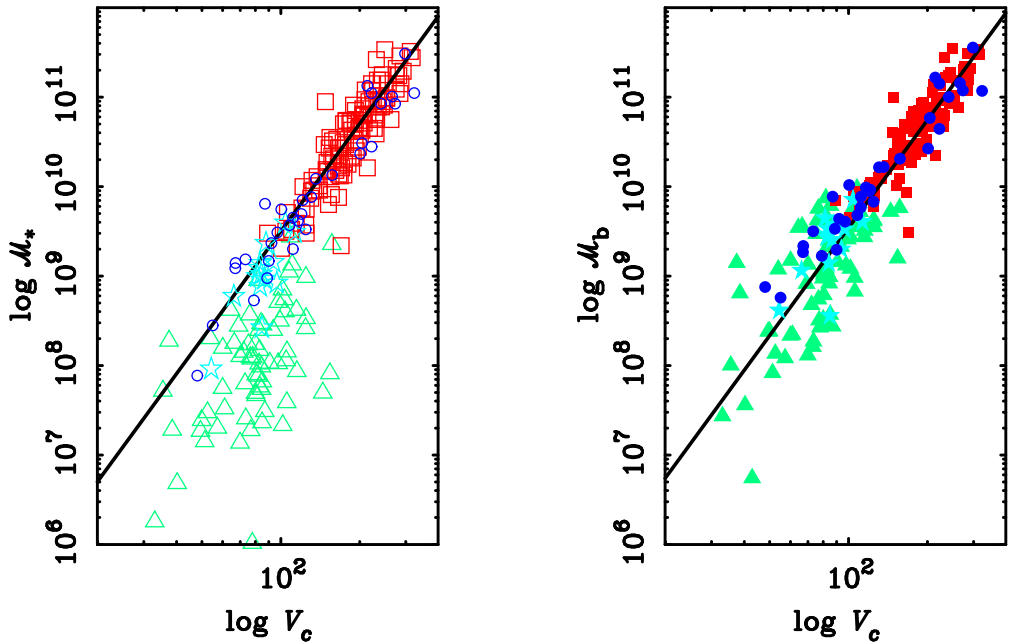}
\caption{The ordinary (left) and baryonic (right) Tully-Fisher relations.
On the left, stellar mass is plotted instead of luminosity assuming
${\cal M}_* = \Upsilon_* L$, where $\Upsilon_*$ is the 
stellar mass-to-light ratio.  On the right, the total baryonic mass is
plotted using ${\cal M}_b = {\cal M}_* + 1.4 {\cal M}_{HI}$.  Different
data sets are represented by different symbols.  Squares:
Bothun et al.\ (1985) $H$-band data with $\Upsilon_*^H = 1.0$.
Circles: McGaugh \& de Blok 1998 $B$-band data with
$\Upsilon_*^B = 1.5$.
Stars: Matthews et al.\ (1998) data (also $B$-band).  Triangles:
Eder \& Schombert (1999) $I$-band data for dwarf galaxies with
$\Upsilon_*^I = 1.5$.  The inclusion of the new dwarf data vastly
increases the dynamic range. The assumed value of $\Upsilon_*$
hardly matters in this log-log plot,
but the inclusion of gas mass for gas rich galaxies certainly does.
There is a clear break in the ordinary Tully-Fisher relation
around $V_c \approx 80$ km s$^{-1}$.  Yet the
Baryonic Tully-Fisher relation appears to be continuous and
linear over nearly 5 decades in mass.
The lines are not fits to the data, but rather have a slope of 4 with
normalization chosen by eye.  
The Tully-Fisher relation appears fundamentally
to be a relation between rotation velocity and luminous baryonic mass
of the form ${\cal M}_b \propto V_c^4$.}
\label{BTF}
\end{figure}

It appears that the Tully-Fisher relation is indeed representative of a
more fundamental relation between total baryonic mass and rotation velocity. 
This is consistent with the simple reasoning underpinning the standard
picture, and would appear to be a success thereof.  Note, however,
that this only follows if
the baryon fraction is constant from galaxy to galaxy and the fraction
$m_d$ of baryons which cool into disks is universal.  While
the first of these is certainly sensible, the second seems difficult to
arrange unless $m_d \rightarrow 1$.  If not, there should be substantial
scatter about the mean $m_d$.  This would propagate into the scatter in
the Tully-Fisher relation whose error budget is already oversubscribed.
However, $m_d = 1$ is not currently expected in CDM models
(e.g., Steinmetz \& Navarro 1999), and it may be necessary to vary $m_d$
in order to fit other constraints (Mo, these proceedings).
In addition, the slope preferred by CDM is
$\sim 3$ (Mo et al.\ 1998) rather than the observed\footnote{The slope of
4 observed for the Baryonic Tully-Fisher relation is also observed in the
$K$-band (Verheijen 1997) and the $I$-band (Sakai et al.\ 1999).} 4.  So
there remain substantial hurdles to understanding the Tully-Fisher
relation in the standard framework.  The rigorous relation between
total luminous mass and rotation velocity illustrated in Fig. 1 provides an
important constraint on theory.

\section{Disk Masses: Maximal or Not?}

There is considerable evidence that the disks of high surface brightness
galaxies are maximal or nearly so (e.g., Sellwood 1999), while those
of low surface brightness galaxies are not (de Blok \& McGaugh 1997).

Maximum disk has, at the least, proven its utility (property 2) in fitting
the inner parts of the observed rotation curves of high surface brightness
galaxies in considerable detail (Fig. 2).  The inference often drawn from
this is that disks are indeed maximal and dark matter is only important at
large radii --- why else would the observed luminous mass so closely
trace the dynamics?  Indeed, if high surface brightness disks are not
maximal, there must be a peculiar conspiracy which causes the
[quasi-spherical] dark matter halo to produce a rotation curve which has the
same shape as that of the luminous disk. This is illustrated in Fig. 2, and
holds regardless of the dark matter fraction since subtracting a
submaximal disk preserves the shape of the rotation curve.  This is
another indication of the strong link between dark and luminous mass over
and above that already implied by the Tully-Fisher relation.

\begin{figure}[t]
\plotone{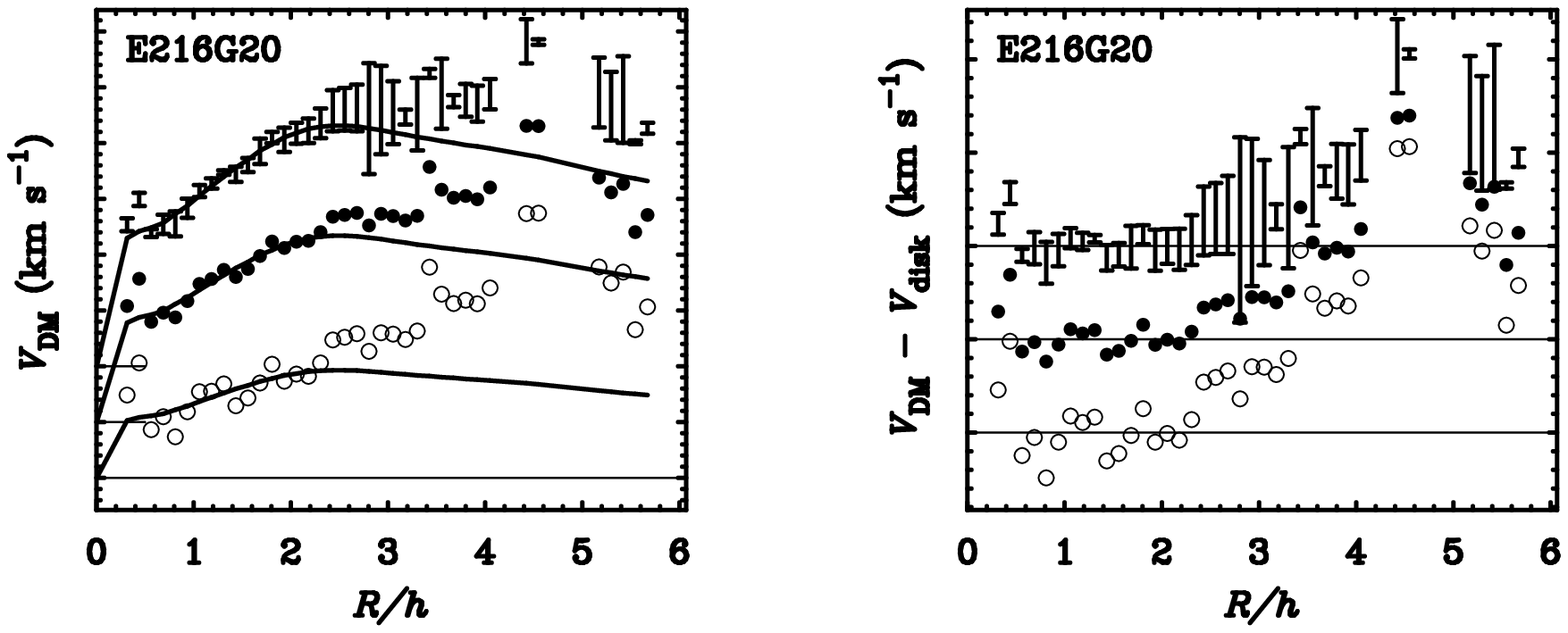}
\caption{The utility of maximum disk fits in bright galaxies is illustrated
by this example.  On the left is plotted the rotation curve due to the
dark matter assuming the disk contributes, at two scale lengths,
80\% of the mass (open circles, effectively maximal); 40\% (solid
circles); and zero mass (points with error bars; the observed rotation). 
Also plotted as lines are the appropriately scaled maximum disk fits (data
and fits from
Palunas 1996).  Each rotation curve is offset by 50 km s$^{-1}$ for clarity.
Since the rotation curve is well described by the shape of the luminous
mass distribution out to $\sim 2.5$ scale lengths, the dark matter must
have a distribution which produces this same shape over this region
irrespective of the amplitude of its contribution.  This is further
illustrated on the right, where the deviation of the dark matter rotation
curve from the shape specified by the disk is shown.}
\label{maxdisk}
\end{figure}

Standing in contrast to this is the large amount of dark matter enclosed
by low surface brightness disks (property 3).  As surface brightness
decreases, so does the amplitude of the disk's contribution to the total
rotation.  Yet the total amplitude of the rotation does not decrease, as
illustrated by the lack of deviation from the Tully-Fisher relation (Fig. 3). 
Consequently, low surface brightness galaxies must be dark matter
dominated.

\begin{figure}[t]
\plotone{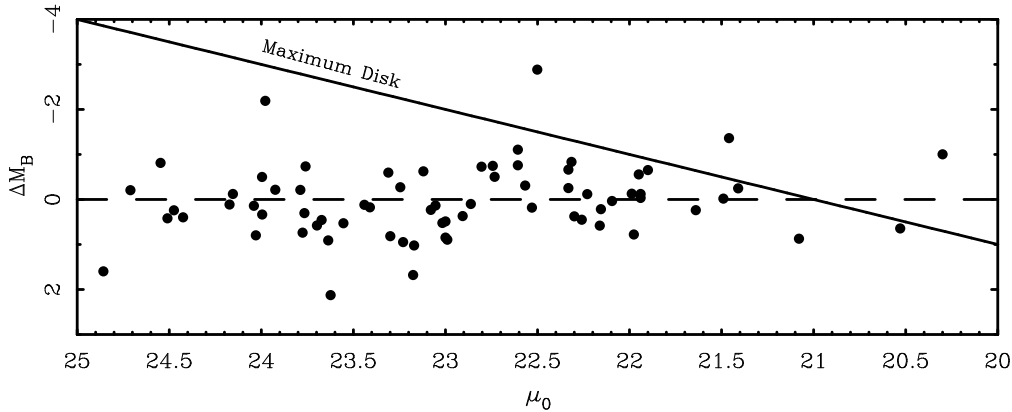}
\caption{Deviations from the Tully-Fisher relation as a function of
surface brightness.  The lack of
systematic deviation is puzzling, since stars have mass and should
contribute some to the measured velocity.  The slope of the deviation
expected for maximal disks is illustrated.}
\label{TFSBresid}
\end{figure}

It is possible that only the highest surface brightness disks are maximal,
or that the utility of maximum disk is a fluke.  On the other hand, it is
sometimes possible to force maximal disk fits to well resolved rotation curves
of low surface brightness galaxies (Swaters, private communication),
albeit at the price of absurdly high stellar mass-to-light ratios.  This just
creates another missing mass problem as we end up invoking dark matter
in the disk rather than in a halo.

\begin{figure}[t]
\plotone{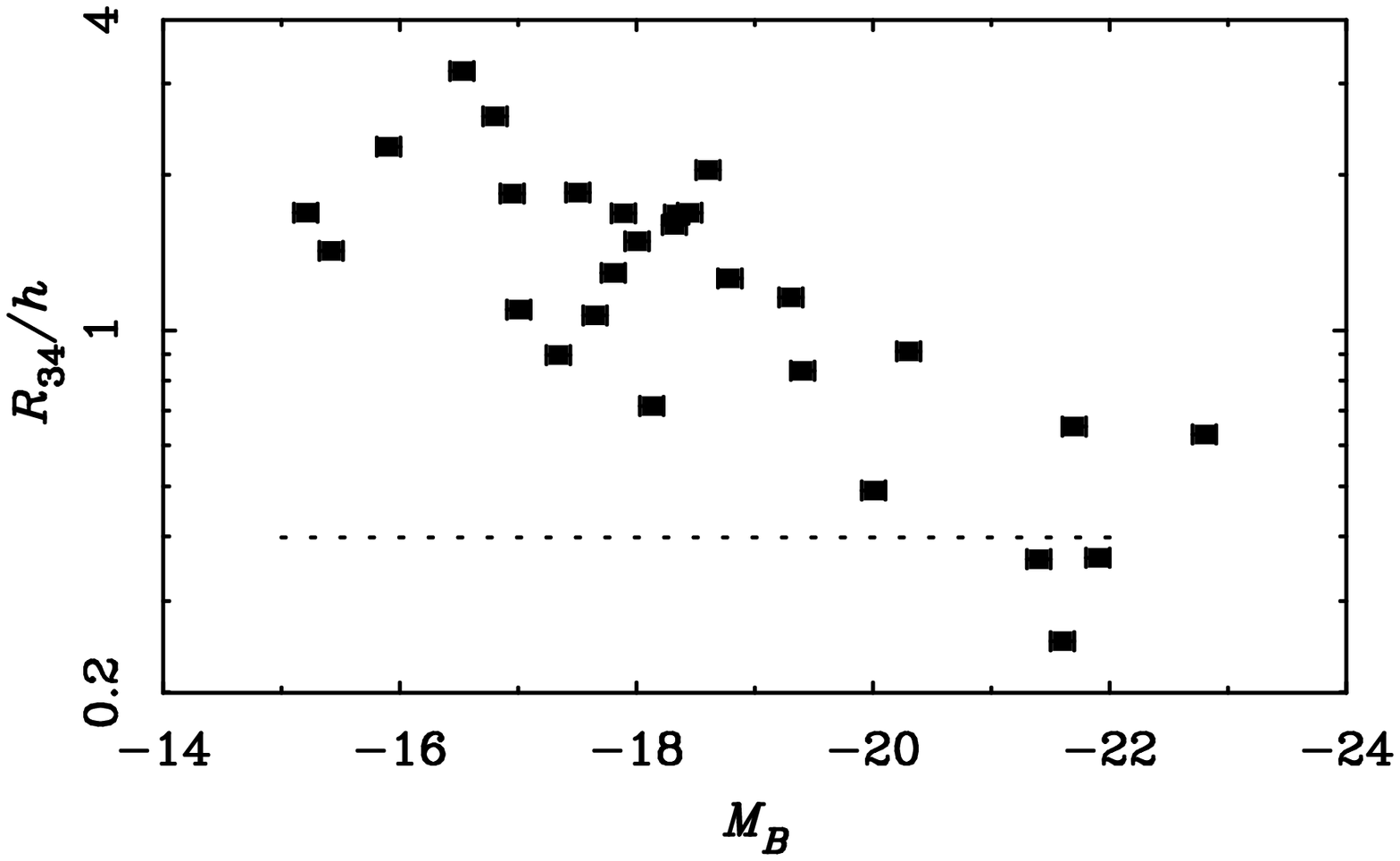}
\caption{The shape of rotation curves is correlated with luminosity.
Here the shape is measured by the radius $R_{34}$ where the rotation
curve has risen to \threequarters\ of its asymptotic value: $V(R_{34}) =
\threequarters V_{flat}$.  This only holds when $R_{34}$ is normalized to
the disk scale length $h$:  the mass distribution is intimately connected
to the light distribution.  The prediction of galaxy formation
models (prior to tweaking) is illustrated by the dotted line
(Dalcanton et al.\ 1997; McGaugh \& de Blok 1998; Mo et al.\ 1998).}
\label{R34}
\end{figure}

A curious extension of the apparent contradiction between (2) and (3) is
the relation between luminosity and rotation curve shape (4).  The shape
of a rotation curve is well predicted with knowledge of the luminosity and
scale length of a galaxy: $V(R/h)$ is pretty much the same\footnote{The
similarity of rotation curve shapes in terms of scale lengths means that
the radius at which a velocity is measured for Tully-Fisher does not matter
much (cf.\ Courteau \& Rix 1999).}
at a given luminosity.  Moreover, rotation curve shape is
well correlated with luminosity.  Bright galaxies have rapidly rising
rotation curves (measured in scale lengths) while faint galaxies have
slowly rising rotation curves.  This provides an important test of galaxy
formation theories, whose {\it a priori\/} predictions fail (Fig. 4).

\section{The Mass Discrepancy--Acceleration Relation}

Another connection between the luminous and dark mass is evinced by the
correlation of the amplitude of the mass discrepancy with centripetal
acceleration (property 5).
For $V_c^2/R > a_{obs} \approx 10^{-10}$ m s$^{-2}$, there is no perceptible
mass discrepancy in spiral galaxies (Sanders 1990).  The luminous stars
and gas suffice to explain the observed rotation for any
reasonable choice of stellar mass-to-light ratio (McGaugh 1999).
For $V_c^2/R < a_{obs}$, the severity of the mass discrepancy increases
systematically with decreasing acceleration.

The fact that there is this acceleration scale where dark matter is needed
provides a constraint on halo parameters
(see also Brada \& Milgrom 1999).  The maximum acceleration
due to the halo, $a_{max}$, can not exceed the observed scale $a_{obs}$.
The absolute value of $a_{obs}$ depends on 
the mass-to-light ratio of the stars.  If $\Upsilon_*$ is low, dark
matter becomes important at slightly higher accelerations, and vice versa.
Since we must guestimate the stellar mass-to-light ratio ($\Upsilon_*^B
=2$ is assumed), let us define the parameter
${\cal Q}$ to be the ratio of assumed to true stellar mass to light ratios
(McGaugh 1999).  This is the one free parameter which can be adjusted to
vary the effects of this limit.

\begin{figure}[t]
\plotone{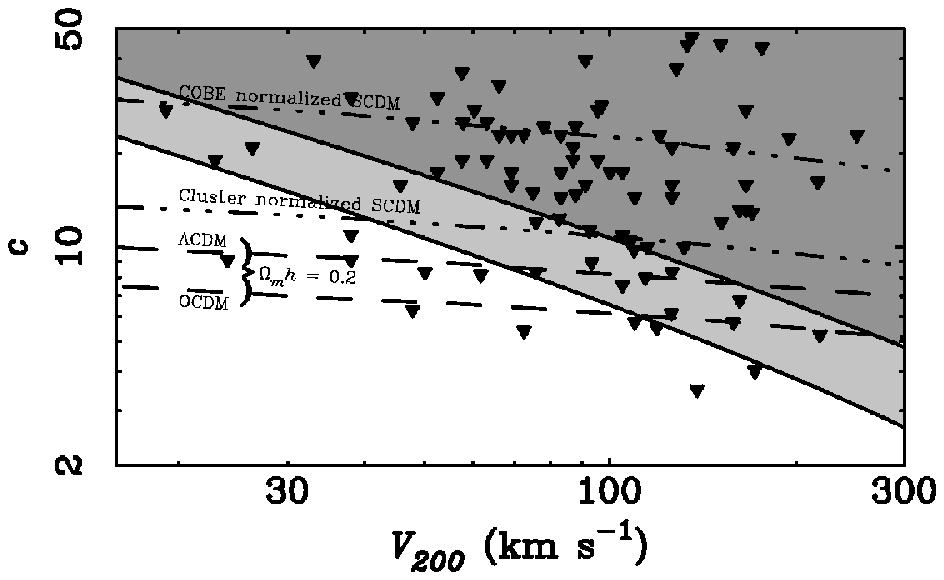}
\caption{The $(V_{200},c)$ parameter space of NFW halos.  Dashed and
dotted lines illustrate the predicted run of halo parameters in particular
cosmologies (as specified).  Triangles indicate upper limits on the
concentration parameter
from fits to high quality galaxy rotation curves.  Only low $\Omega_m h$
cosmologies appear to be viable.  The shaded regions illustrate the region of
parameter space excluded by the acceleration scale ($a_{max} > a_{obs}$)
for ${\cal Q} = 1$ (light gray) and ${\cal Q} = 4$ (dark gray).}
\label{NFWcV}
\end{figure}

As an example of the utility of this constraint, consider the form of dark
matter halos found in numerical simulations of structure formation (for
example, Navarro, Frenk, \& White 1997; NFW).
The maximum acceleration of an NFW halo occurs at $R = 0$.
From the NFW expression for circular velocity, the acceleration limit is
\begin{equation}
a_{max} = \frac{c^2 V_{200}}{\ln(1+c)-c(1+c)^{-1}} < {\cal Q}^{1/2}
a_{obs}.
\end{equation}
This is shown in Figure 5 for ${\cal Q} =1$ and 4 corresponding to
true $\Upsilon_*^B = 2$ and \onehalf\ respectively.
Quite a lot of parameter space is disallowed unless we are prepared to
consider $\Upsilon_* \ll \onehalf$.

\section{The MOND Phenomenology}

There are two ways of fitting the observed rotation curves of spiral
galaxies.  In the context of dark matter, we subtract the observed
luminous mass and attribute the remainder to dark matter.  Thus
\begin{equation}
V_c^2(R) = V_*^2(R) + V_g^2(R) + V_h^2(R),
\end{equation}
where $V_c$ is the total rotation, $V_*$ is the component due to the
stars, $V_g$ is that due to the gas, and $V_h$ is that of the halo.
An equally successful procedure is offered by MOND, which has no dark
matter.  The Newtonian contributions of the luminous components are
modified so that
\begin{equation}
g_N = \mu(x) \frac{V_c^2(R)}{R},
\end{equation}
where $x = a/a_0 = V_c^2/(a_0 R)$ and $\mu(x)$ is an arbitrary
interpolation function with the asymptotic properties
$\mu(x) \rightarrow 1$ for $x \gg 1$ and
$\mu(x) \rightarrow x$ for $x \ll 1$.  The total rotation in MOND is thus
related to the Newtonian contributions of the luminous components
through
\begin{equation}
V_c^2(R) = \mu^{-1}(x) [V_*^2(R)+V_g^2(R)].
\end{equation}

We can equate the two ($V_c = V_c$), and solve for the single variable
which is not directly observed, $V_h$:
\begin{equation}
V_h^2(R) = \mu^{-1}(x)[{\cal Q} V_*^2(R) + V_g^2(R)]
\end{equation}
It is thus possible to uniquely express the mass distribution of the dark
matter
halo in terms of that of the luminous matter and a simple function of the
observed centripetal acceleration.  The only free parameter is ${\cal Q}$,
which allows for a difference between the true mass-to-light ratio and
that required by MOND.

The fact that MOND provides a phenomenological description of rotation
curves imposes a very strong requirement on galaxy formation theories.
If a theory is really to describe
reality, it must predict the detailed distribution of all three
mass components:  stars, gas, and dark matter.  One must then show
that the predicted distributions satisfy the MOND phenomenology: it
should be possible to obtain a good MOND fit to the model galaxy
using only the luminous components.  The only free parameter with
which to do this is ${\cal Q}$.  There is no fudge factor associated with
the gas distribution, and the freedom in the stars with ${\cal Q}$ is very
limited since MOND returns very reasonable mass-to-light ratios
for stellar populations.  This means ${\cal Q} \approx 1$.  
Dark and luminous mass must be very strongly coupled indeed!

\section{Conclusions}

There now exists a lengthy list of dynamical properties constraining
theories of disk galaxy formation.  These properties indicate a remarkably
strong link between luminous and dark matter.  No extant galaxy formation
model passes all, or even most, of the tests posed by the observed
properties.  Relatively straightforward models (e.g., Dalcanton et al.\
1997; McGaugh \& de Blok 1998; Mo et al.\ 1998) require extensive tweaking;
so far only baby steps
have been taken in this direction (van den Bosch, these proceedings).

\acknowledgments I am grateful to Erwin de Blok, Vera Rubin, Simon White,
Houjun Mo, James Binney, Jerry Sellwood, Renzo Sancisi, Povilas Palunas,
and Rob Swaters for conversations relating to the topics discussed here.

\end{document}